\pdfoutput=1
\newcommand{\magcir}{\raise -2.truept\hbox{\rlap{\hbox{$\sim$}}\raise5.truept\hbox{$>$}\ }}
\newcommand{\etal}{{\it et al. }}
\def\lsimeq{{_<\atop^{\sim}}}
\def\gsimeq{{_>\atop^{\sim}}}

\documentclass{PoS}

\title{The AGN Component in Deep Radio Fields: Current Understanding}

\ShortTitle{Understanding The AGN Component in Deep Radio Fields}

\author{\speaker{Isabella Prandoni}\\
        INAF - Istituto di Radioastronomia, Via P. Gobetti 101, 40129 Bologna, Italy\\
        E-mail: \email{prandoni@ira.inaf.it}}

\abstract{The present paper reviews our current understanding of the AGN 
component in sub--mJy radio fields, as it results from the exploitation of 
multi-frequency information available in two deep extra-galactic radio fields: the ATESP 5~GHz sample and the First Look Survey.
One of the key issues addressed here is whether low-power AGNs are more related to efficiently accreting systems (mostly 
radio-quiet) or to systems with very low accretion rates (mostly radio-loud). 
The emerging picture is the following. Radio-loud jet-dominated radio galaxies  seem 
to be largely dominant  down to flux densities of the order of e.g. $S>400$ $\mu$Jy.
At lower  flux densities ($S_{\rm 1.4 GHz}\magcir 100$ $\mu$Jy) radio-loud AGN are still present in significant numbers. However 
a population of radio--emitting AGNs, whose properties are consistent with those 
expected from existing radio--quiet AGN modeling, clearly shows up. This may indicate that the bulk of the radio--quiet 
AGN population could emerge from studies of deeper ($S<100$ $\mu$Jy) radio samples.
The radio-quiet AGN component could be recognised thanks to the availability 
of IR colors which prove to be especially useful to efficiently separate radio sources triggered by AGNs, from sources 
triggered by star-formation. }

\FullConference{ISKAF2010 Science Meeting - ISKAF2010\\
		June 10-14, 2010\\
		Assen, the Netherlands}

\begin{document}

\section{Introduction}
\begin{figure}
\begin{center}
\includegraphics[width=.7\textwidth]{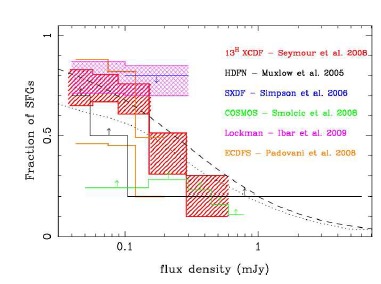}
\caption[]{Fraction of star-forming galaxies as a function of flux density in several recently studied deep radio fields, as summarized by Seymour~\etal (\cite{Sey08}). 
\label{fig:seymour}}
\end{center}
\end{figure}

\begin{figure}
\begin{center}
\includegraphics[width=.7\textwidth]{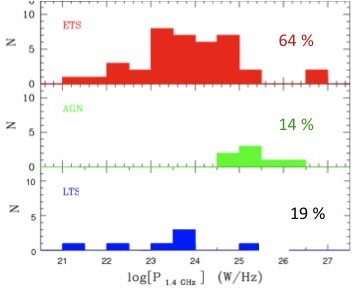}
\caption[]{Distribution of the ATESP radio sources studied by Mignano~\etal (\cite{Mig08}) as a function of 1.4 GHz radio power. Red histogram: early-type galaxies. Green histogram: broad-line AGNs/QSO. Blue histogram: late-type/star-forming galaxies.
\label{fig:phist}}
\end{center}
\end{figure}

After many years of studies, AGNs are now recognised to contribute 
significantly at radio fluxes below 1 milliJy (mJy).\\
Multi-wavelength studies of deep radio fields show that star-forming
galaxies dominate at microJy ($\mu$Jy) levels, and rapidly decrease in number
going to higher flux densities. This is very clearly shown by Seymour~\etal (\cite{Sey08}), who 
graphically summarizes the results obtained by several recent studies in a single self-explanatory 
picture (see Fig.~\ref{fig:seymour}). From the analysis reported in Fig.~\ref{fig:seymour} it is clear that at 
$S>100-200$ $\mu$Jy star-forming galaxies account for no more 
than $20-30\%$ of the entire population. At such flux densities the sub-mJy 
population is in fact dominated by AGNs. This is very clear in Fig.~\ref{fig:phist}, which shows the composition 
of the ATESP 5~GHz sample, characterized by a flux limit of 400 $\mu$Jy (see \cite{Pra06, Mig08}).
As reported by Mignano~\etal (\cite{Mig08}), radio sources associated with early--type galaxies and plausibly triggered by AGNs 
account for  64\% of the total (red histogram) with a further 14\% contribution from broad-/narrow-line AGNs (green histogram). 
Star-forming galaxies (blue histogram) account for only 19\%. 
Evidences pointing towards a large AGN contribution at sub-mJy fluxes have been found in several other studies of deep radio fields
(see e.g. \cite{Gru99, Geo99, Mag00, Pra01b, Mux05, Afo06, Smo08, Pad09}).\\

This somehow unexpected presence of large numbers of AGN--type sources at 
sub-mJy levels has given a new and interesting scientific perspective to the
study of deep radio fields, since a better 
understanding of the physical and evolutionary properties of 
low/intermediate power AGNs may have important implications for the 
determination of the black-hole-accretion history
of the Universe as derived from radio-selected samples. 
Of particular interest is the possibility of assessing whether the AGN
component of the sub--mJy population is more related to
efficiently accreting systems - like radio-intermediate/quiet quasars - or to
systems with very low accretion rates - like e.g. Fanaroff \& Riley Type I (FRI) radio galaxies 
(\cite{Fan74}). The
latter scenario (radio mode) is supported by the presence of many optically
inactive early type galaxies among the sub--mJy radio sources. The
quasar mode scenario, on the other hand, may be supported by the large 
number of so-called
radio-intermediate quasars observed at mJy levels by e.g. Lacy \etal
(\cite{Lac01}) and by the modelling work of Jarvis \& 
Rawlings (\cite{Jar04}), 
who predict a significant contribution of radio-quiet quasars ($P\sim 10^{22-24}$ W Hz$^{-1}$), at 
sub-mJy levels. Radio-quiet AGNs are expected to be compact and have steep radio spectrum (\cite{Kuk98}).\\
The sub-mJy radio population is currently modeled in three main components. The star-forming galaxies, the extrapolation to low flux densities of the classical radio-loud AGN population (radio galaxies and radio-QSO, which fully account for the source counts at higher flux densities), and a radio-quiet AGN component (assuming they are not radio silent). \\
Another issue is represented by the possible role played at sub-mJy radio fluxes by low 
radiation efficiency accretion mechanisms, associated to optically thin discs,
such as the so-called {\it advection dominated accretion flows} (ADAF) and 
modifications (ADIOS, CDAF, etc; see \cite{Nar94, Qua99, Abr02}).\\
In order to address the physical properties and nature of the sub-mJy AGN component, I will focus 
in the following on recent results obtained from multi-wavelength studies of two deep extra-galactic radio fields: the already mentioned ATESP 5~GHz survey and the First Look Survey (FLS).

\section{The AGN Component in the ATESP 5~GHz Sample}
\begin{figure}
\begin{center}
\includegraphics[width=.8\textwidth]{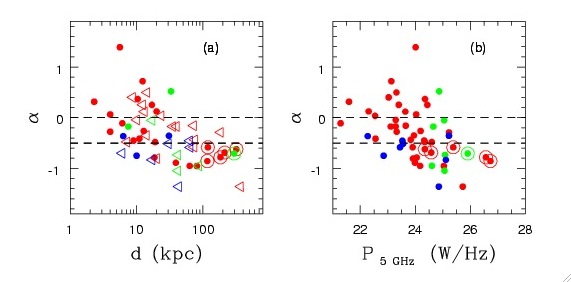}
\caption[]{1.4 - 5 GHz spectral index against source size (panel {\it a})  and 5~GHz radio power (panel {\it b}) for
the sources in the ATESP 5 GHz sample. Color code as  in Fig.~\ref{fig:phist}. Circled dots indicate double/multi-component 
radio sources.
\label{fig:alpha_dim_power}}
\end{center}
\end{figure}

The ATESP 5 GHz survey is especially suited to study the phenomenon of low-luminosity 
nuclear activity, possibly related to low radiation/accretion processes and/or 
radio-intermediate/quiet QSOs, since it is characterized by a relatively high flux density limit 
($S\sim 400$ $\mu$Jy). At such flux densities
star-forming galaxies are starting to appear, but are not yet the dominant population 
(see Fig.~\ref{fig:phist}). The ATESP 5~GHz survey, 
carried out with the Australia Telescope Compact Array (ATCA) by
Prandoni~\etal (\cite{Pra06}), covers 
a $2\times 0.5$ sq.~degr. region of the wider original 1.4 GHz ATESP 
survey (\cite{Pra00a,Pra00b,Pra01a}), and has deep ($R<25$) 
UBVRIJK multi-colour imaging available (\cite{Hil06, Ols06}). 
Interestingly, the analysis of the 1.4 and 5~GHz ATESP data has revealed 
a significant flattening of the source radio spectra going from mJy to 
sub-mJy flux levels (\cite{Pra06}). Such flattening is mostly
associated to early-type galaxies. As shown in 
Fig.~\ref{fig:alpha_dim_power} panel {\it a},
$>60\% $ of the early-type galaxies (red points)  have flat ($\alpha >-0.5$, where $S\sim \nu^{\alpha}$) 
and/or inverted ($\alpha >0$) spectra and rather compact linear sizes ($d<10-30$ kpc, see also \cite{Mig08}).
The absence of emission lines together with low 
radio luminosities (typically $10^{22-25}$ W/Hz, see Fig.~\ref{fig:alpha_dim_power}, panel {\it b}) 
suggests that these are FRI radio galaxies. However, Both compactness and spectral shape 
suggest a core emission with strong synchrotron or free-free self-absorption. 
Such sources are known to exist among FRI radio galaxies, but 
they are relatively rare. 

Alternatively, these sources may represent a composite class of objects very 
similar to the so-called low power ($P_{408 MHz}<10^{25.5}$ W/Hz) 
compact ($<10$ kpc) - LPC - radio sources studied by 
Giroletti~\etal (\cite{Gir05}). LPC host galaxies do not show signatures of
strong nuclear activity in the optical (and X-ray) bands, and preliminary 
results indicate that multiple causes can produce LPC sources: geometrical-relativistic
effects (low power BL-Lacertae objects), youth (GPS-like sources), 
instabilities in the jets, frustration by a denser than average ISM and a 
premature end of nuclear 
activity (sources characterised by low accretion/radiative efficiency, i.e. 
ADAF/ADIOS systems). However some difficulties remain. First, very simple orientation
arguments show that no more than 3-6\% of the sources in the sample can be associated to BL-Lac 
sources (about $2-4$ objects assuming viewing angles $<15^{\rm o}-20^{\rm o}$).
Second, if the 
objects are young GPS sources, much higher ($>10^{25}$ W/Hz)
radio luminosities are expected.
Third, if they are old ADAF/ADIOS sources, the expected luminosities are
much lower ($<10 ^{21}$ W/Hz, \cite{Doi05}). 
In the latter case there may yet be different solutions, as for example if 
the ADAF source coexists with a radio jet, as was suggested by Doi \etal (\cite{Doi05}) 
in the case of a few low luminosity AGNs (see also the ADAF-jet model described in
\cite{Fal99}).\\
Another possibility is that we are dealing with 
radio-quiet/intermediate QSOs, where the activity in the optical band is 
obscured by dust in the galaxy. This scenario may be supported by 
the results reported by Kl\"ockner~\etal (\cite{Klo09}), who have recently 
observed with the EVN 11 $z>2$ radio-intermediate obscured quasars, detecting seven of them. 
The detected radio emission 
accounts for 30-100\% of the entire source flux
density, and the physical extent of this emission is $\lsimeq 150$ pc. 
The missing flux implies the likely existence of radio jets of physical sizes 
between $\gsimeq 150$ pc and $\lsimeq 40$ kpc. 
It is worth noticing on the other hand, that broad-line AGNs and QSO (green points in Fig.~\ref{fig:alpha_dim_power}, panel {\it b}) have radio powers  of the order of $P\sim 10^{25-26}$ W Hz$^{-1}$, higher than expected for radio-quiet QSO.\\

\begin{figure}
\begin{center}
\includegraphics[width=.7\textwidth]{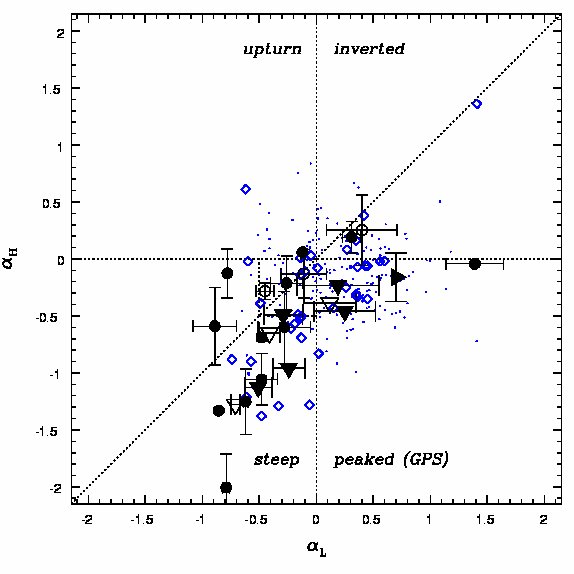}
\caption[]{Color-color radio plot. Black symbols indicate ATESP sources, triangles indicate upper/lower 
limits in $\alpha_H/\alpha_L$ respectively. Blue diamonds indicate sources 
associated with early-type optical spectra from  the Massardi \etal sample (\cite{Mas08}; 
blue dots indicate sources associated with quasars or quasar 
candidates in the Massardi \etal sample. 
\label{fig:alphaplot}}
\end{center}
\end{figure}

To further explore the nature of the AGN component in the ATESP 5 GHz sample,  multi-frequency
(4.8, 8.6 and 19 GHz) quasi-simultaneous observations of a complete sub-sample of ATESP radio sources associated 
with early-type galaxies (26 objects with $S>0.6$ mJy) was carried out with the ATCA (\cite{Pra10}). This mutli-frequency data can 
provide insights into the accretion/radiative mechanism that is at work, since different regimes display different spectral signatures in 
the radio domain.\\
Figure~\ref{fig:alphaplot} shows the so-called color-color radio plot: high frequency (5/8 to 20 GHz) spectral index against the low 
frequency (1.4 to 5 GHz) one. The sources (black symbols) preferentially populate the regiom below the diagonal line, indicating a mild steepening of their spectra going from lower to higher radio frequencies. More importantly,  all sources have high frequency spectral index $<0.2$, i.e. consistent with jet-dominated surces, with the flattest sources possibly having a major contribution from base self-absorbed jet components. Pure ADAF models predict strongly inverted spectra and are ruled out by the high frequency data, while ADAF+jet scenarios are still consistent with flat/moderately inverted-spectrum sources, but are not required to explain the data. \\
A comparison with the {\it Bright Source Sample} ($S>500$ mJy), extracted by Massardi \etal (\cite{Mas08}) from the 20 GHz AT20G survey,
shows that the radio spectral properties of the ATESP early-type sources are more similar to the ones of brighter radio galaxies (blue diamonds), than to the ones or brighter radio-quasars (blue dots). This support the hypothesis that at the flux densities probed by the ATESP sample ($S>400$ $\mu$Jy), the AGN component is mainly the low power extrapolation of the classical bright radio galaxy population, rather than the radio-quiet counterpart of bright radio quasars. 

\section{Looking for a radio-quiet AGN component in the First Look Survey}
In order to probe the radio-quiet AGN component it seems necessary to analyze deeper radio fields, like for instance the 
First Look Survey ($S>100$ $\mu$Jy). 
As part of the extragalactic component of the First Look Survey (FLS), 
a region covering 4 square degrees and centered on RA=17:18:00, 
DEC=59:30:00 was imaged, with the aim of studying a low Galactic Background 
region to a significantly deeper level than any previous large-area 
extragalactic infrared survey. 
Such a survey was complemented by a smaller ~0.75x0.3 sq. degr. 
survey ('verification' survey), lying in the same region, observed 
to a factor of 3 deeper flux levels. \\
Spitzer images and source catalogues are available at 3.6, 4.5, 5.8, 8.0 
$\mu$m (IRAC, \cite{Lac05}) and at 24, 70, 160 $\mu$m (MIPS, 
\cite{Fad06,Fra06}), complemented by a large set of ancillary data taken 
at different wave-bands: from deep optical imaging (R-band, \cite{Fad04}; 
u*-/g*-bands, \cite{Shi06}) and spectroscopy (\cite{Sto02,Pap06,Mar07})
to deep (rms noise level~23-30 muJy)
radio images at 1.4 GHz (VLA, \cite{Con03}) and 610 MHz (GMRT, \cite{Gar07}). 
A deeper (rms noise level ~8.5 muJy) 1.4 GHz mosaic was obtained 
at Westerbork for a 1~sq.~degr. region covering the FLSv (\cite{Mor04}). \\
The availability of both deep radio and far-infrared data is of particular 
interest, since it is possible to exploit the well-known tight correlation 
between far-IR and radio luminosities of star-forming galaxies
(see e.g. \cite{Hel85,Gar02})
to efficiently separate radio sources triggered by AGNs. 
Very useful is also the availability of data at two radio frequencies -- 0.61 
and 1.4 GHz -- which allows us to derive the source spectral index ($\alpha$), as done for the ATESP sources.

\begin{figure}
\begin{center}
\includegraphics[width=.8\textwidth]{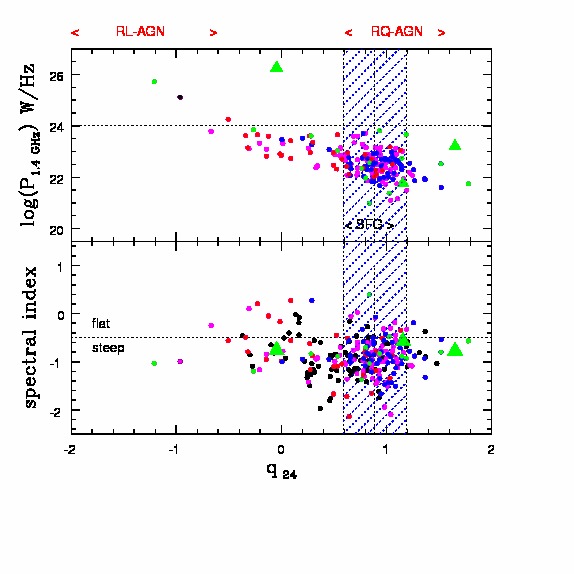}

\vspace{-1.5cm}
\caption{1.4 GHz luminosity (top) and $0.61-1.4$ GHz spectral index (bottom) 
as 
a  function of $q_{24}=log(S_{24\mu m}/S_{1.4GHz})$ for the optically identified 
FLS radio sources. 
The  expected location or star-forming galaxies is shown by the blue 
shaded region. Symbols
refer to optical spectral classification: star-forming galaxies (blue); 
early-type galaxies (red); broad/narrow emission-line AGN spectra (green 
triangles/dots); spectra showing narrow emission lines, but whose 
spectral features 
do not allow an unambiguous SFG vs. AGN classification (magenta); no spectral 
information (black). From Prandoni \etal (\cite{Pra09}).
\label{fig:q24plot}}
\end{center}
\end{figure}

The FLS 0.61-1.4 GHz radio catalogue was cross-correlated with the 
Spitzer IRAC multi-color (3.6, 4.5, 5.8, 8.0 $\mu$m) catalogue (\cite{Lac05}), 
with the MIPS 24 $\mu$m catalogue (\cite{Fad06}) and with the 
optical spectroscopy catalogues of \cite{Mar07} and \cite{Pap06}. 
Identifying the optical counterpart of the 
sources is crucial to get information on both the galaxy redshift and 
classification (broad/narrow-line AGN, star-forming or early-type galaxy). 
An optical identification was 
found for $\sim 20\%$ of the sources, most of which have a measured redshift. 
The sample of optically identified sources was then used to extract 
a robust sub-sample of radio-emitting AGNs.\\
A first analysis of the multi-wavelength properties of the optically 
identified 
radio sources in the FLS is illustrated in Figure~\ref{fig:q24plot}, where 
the 1.4 GHz radio power (top) and the 0.61-1.4 GHz spectral index 
(bottom) are plotted as a function of the so-called $q_{24}$ parameter, defined as 
the ratio between the 24 $\mu$m and the 1.4 GHz source flux density 
($q_{24}=log(S_{\rm 24mu m}/S_{\rm 1.4GHz}$). The $q_{24}$ value range allowed for 
star-forming galaxies is shown by the blue shaded region (\cite{Mar07}). 
Radio-loud AGNs are located at the left side of such region (corresponding to 
an excess of radio emission with respect to infrared). 
As found in 'brighter' sub-mJy samples (see e.g. the ATESP sample, 
\cite{Mig08}) we have a significant fraction of radio-loud AGNs in the FLS, 
and among them we have several flat-/inverted--spectrum sources. 
Such radio--loud AGNs mostly have low radio
powers ($P_{\rm 1.4 GHz}<10^{24}$ W/Hz) and are preferentially identified with 
early-type galaxies (shown in red) or weak narrow line systems (shown in 
magenta). Such properties are consistent with those of low power (FRI-type) 
radio galaxies, typically characterized by very low accretion rates. 

\begin{figure}
\begin{center}
\includegraphics[width=.8\textwidth]{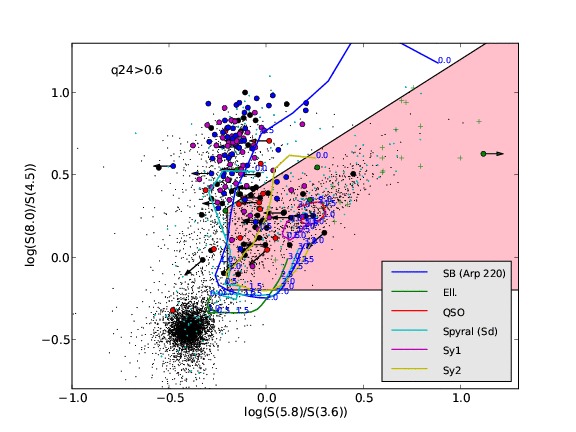}
\caption[]{IRAC color-color plot of the FLS radio sources with $q_{24}>0.6$ 
(filled symbols), i.e. for sources with infrared-to-radio ratios consistent 
with the ones of star-forming galaxies and/or radio-quiet AGNs.
Colors refer to optical spectral classification as in Figure~\ref{fig:q24plot}.
Arrows indicate upper/lower limits. 
The expected IRAC colors as a function of redshift for different source types 
are shown by different lines (see legenda in the plot). The expected
location for AGNs is highlighted in pink.
For reference we also show IRAC colors of: {\it a)} all FLS IRAC-identified 
radio sources (no optical identification selection applied, cyan dots); 
{\it b)} the entire FLS IR-selected star/galaxy population 
(no radio selection applied, black dots); and {\it c)} a sample of  
high redshift obscured (type-2) quasars 
(see Martinez-Sansigre et al. 2006, green crosses). From Prandoni \etal (\cite{Pra09}).
\label{fig:IRACplot}}
\end{center}
\end{figure}

As shown in Figure~\ref{fig:q24plot} the bulk of the FLS radio sources are 
characterized by $q_{24}>0.6$, i.e. by $q_{24}$ values
consistent with the sources being star-forming galaxies.  
Nevertheless only a fraction of such objects is optically classified as 
star-forming galaxy (blue points). A few sources are instead optically 
classified as AGNs (green points). Such AGNs represent
a first very clear direct evidence of a radio-quiet AGN population 
showing up at the radio flux levels of the FLS. 
In addition there are a few sources classified as 
early-type galaxies (red points) and a significant fraction of sources 
displaying narrow emission lines, which do not have a secure classification 
(magenta points). Among such sources there may be several other hidden 
radio--quiet AGNs.\\
In order to better disentangle radio sources triggered by star 
formation from those triggered by AGNs, the available 
IRAC colors were exploited. Figure~\ref{fig:IRACplot} shows the IRAC color-color plot for
FLS sources with $q_{24}>0.6$. As expected sources 
optically classified as star-forming galaxies (blue points) are confirmed as 
such by their IRAC colors, together with many of the sources with no secure 
optical classification (magenta points). Nevertheless there is a fraction of 
sources with typical AGN IRAC colors (objects falling in the pink region), 
which can be considered as genuine radio-quiet AGNs. Such radio-quiet AGNs account for $\sim 45\%$ of the overall
AGN component in the FLS.
Their radio properties ($P_{\rm 1.4 GHz}< 10^{24}$ W/Hz; steep radio spectrum, 
see Figure~\ref{fig:q24plot}) 
are fully consistent with those expected for the radio-quiet AGN population 
and, as expected, they are mostly associated to galaxies showing  narrow emission 
lines in their optical spectra (\cite{Jar04,Kuk98}). 

\section{Conclusions and Future Perspectives}
From the analysis of the ATESP 5~GHz sample and of the First Look Survey we can draw the following conclusions:

\begin{itemize}
\item down to  flux densities $S\sim 400$ $\mu$Jy there is no clear evidence for a radio-quiet AGN component;
\item sub-mJy radio-loud AGNs are mostly low-power, compact, (self-absorbed) jet-dominated systems, and seem to be mostly consistent 
with being a low power counterpart of the classical bright radio galaxy population;
\item going to lower flux densities radio-quiet AGNs become increasingly important and at  $S>100$ $\mu$Jy they
account for about 45\% of the overall AGN component;
\item mid-IR (IRAC) color-color diagrams prove to be very powerful to distinguish star-forming galaxies from radio-quiet AGNs;
\item radio-quiet AGNs show radio/optical/IR properties consistent with the ones expected from available modeling and radio follow-up of optically selected radio-quiet AGNs: compact radio sizes; steep radio spectra; $P<10^{24}$ W Hz$^{-1}$; Sy 2 optical spectra and/or mid-IR colors.
\end{itemize}

\begin{figure}
\begin{center}
\includegraphics[width=.7\textwidth]{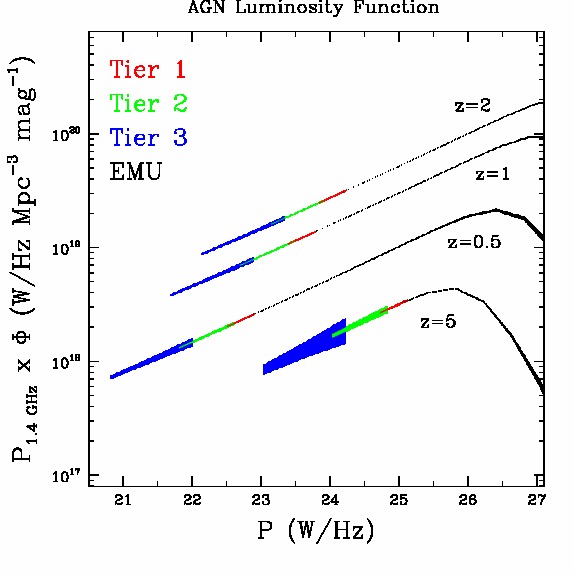}
\caption[]{AGN luminosity function at different redshifts that can be derived putting together ASKAP and MeerKat deep surveys. 
\label{fig:flum}}
\end{center}
\end{figure}

It is clear however that panchromatic analyses of larger and deeper radio samples are needed to confirm 
the above results on a more reliable statistical basis, and to obtain robust quantitative constraints to current evolutionary modeling of faint radio-loud and radio-quiet AGNs . 
To this respect a huge step forward will be done in the next years, thanks to the combination of wide-field/all-sky surveys (sampling large local volumes) and very deep fields (sampling low powers at  high redshifts), planned with the upcoming facilities (LOFAR, ASKAP, MeerKat, etc.).
An example of what can be obtained is illustrated in Fig.~\ref{fig:flum}, where we show the AGN luminosity function at different redshifts that can be derived from the combination of the planned ASKAP all-sky survey ({\it EMU}) and of the three-tiered deep survey proposed for MeerKat.  

\end{document}